# Tinselenidene: a Two-dimensional Auxetic Material with Ultralow Lattice Thermal Conductivity and Ultrahigh Hole Mobility

Li-Chuan Zhang,[†] Guangzhao Qin,[†] Wu-Zhang Fang,[‡] Hui-Juan Cui,[‡] Qing-Rong Zheng,[‡] Qing-Bo Yan,[*,†] Gang Su[*,‡]

†College of Materials Science and Opto-Electronic Technology, University of Chinese Academy of Sciences, Beijing, China 100049,

‡School of Physical Sciences, University of Chinese Academy of Sciences, Beijing, China 100049

**Corresponding Author**

*Email: (Q.-B.Y.) yan@ucas.ac.cn.

*Email: (G.S.) gsu@ucas.ac.cn.

# ABSTRACT

By means of extensive ab initio calculations, a new two-dimensional (2D) atomic material tin selenide monolayer (coined as tinselenidene) is predicted to be a semiconductor with an indirect gap (~1.45 eV) and a high hole mobility (of order 10000 $cm^2V^{-1}S^{-1}$), and will bear an indirect-direct gap transition under a rather low strain (< 0.5 GPa). Tinselenidene has a very small Young's modulus (20-40 GPa) and an ultralow lattice thermal conductivity (< 3 $Wm^{-1}K^{-1}$ at 300 K), making it probably the most flexible and most heat-insulating material in known 2D atomic materials. In addition, tinseleniden has a large negative Poisson's ratio of -0.17, thus could act as a 2D auxetic material. With these intriguing properties, tinselenidene could have wide potential applications in thermoelectrics, nanomechanics and optoelectronics.

## Introduction

The discovery of graphene leads to an upsurge in exploring two-dimensional (2D) materials,[1] such as hexagonal boron nitride[2], transition metal dichalcogenides,[3] silicone,[4] and others.[5] Recently, few-layer black phosphorus (phosphorene) has been successfully exfoliated,[6,7] arousing wide interest for researchers.[8-17] Phosphorene was found to be a 2D anisotropic semiconductor with a moderate direct bandgap and a high carrier mobility,[18-25] leading to phosphorene a promising candidate in nanoelectronics. Bulk tin selenide (SnSe) [26-29] has a hinge-like layered structure similar to black phosphorous, and have applications in photovoltaic[26] and thermoelectric devices.[27,28] Very recently, 2D SnSe has been synthesized,[29,30] which is expected to be useful with great potential in photodetector and photovoltaic devices. However, the physical properties of SnSe monolayer (coined as tinselenidene), in particular the intrinsic carrier mobility, lattice thermal conductivity, strain effects, etc., are still less known.

Here we systematically investigated the geometrical, mechanical, electronic properties of tinselenidene by utilizing the density functional theory (DFT) based *ab initio* calculations. We found that tinselenidene is a semiconductor with an indirect bandgap of 1.45 eV, a very low lattice thermal conductivity (LTC) below 3 $Wm^{-1}K^{-1}$, a large negative Poisson's ratio of -0.17, and a hole mobility as high as 11000 $cm^2V^{-1}S^{-1}$. In contrast to phosphorene, which was reported to have strong anisotropic mechanical, electronic, and optical properties,[18-20,31] tinselenidene was found to bear nearly symmetric phonon and electronic band structures, and other isotropic properties such as LTC and effective mass of charge carrier, etc. These unexpected isotropic features can be attributed to its effective symmetric bilayer square-like lattice structure. The geometric, mechanic and electronic properties of tinselenidene are sensitive to a strain. The Young's elastic modulus are 24.3 GPa and 43.5 GPa along the armchair and zigzag direction, respectively, which may be the most flexible in known 2D atomic materials. A very low stress (1.6 GPa) along armchair direction can induce a geometrical phase transition. Besides, a uniaxial strain will shift the extremes of different energy valleys asynchronously, giving rise to an indirect-direct bandgap transition under a rather low stress (< 0.5 GPa). Although the effective mass of charge carrier is nearly isotropic, the carrier mobility is highly anisotropic, which can be attributed to the anisotropic response of electronic structure to a strain. With these fascinating properties, tinselenidene could have wide potential applications in thermoelectrics, nanomechanics and optoelectronics.

**Results and Discussion**

Tinselenidene and phosphorene are isoelectronic if only valence electrons are considered, and have similar hinge-like quasi-2D structures (Fig. 1). This structure can be viewed as a deformed honeycomb structure of graphene, and is distinctly different along armchair ($x$) and zigzag ($y$) directions, which may be responsible for the strong anisotropic properties of phosphorene.[18-20,31] Owing to structural similarity, it is interesting to compare tinselenidene and phosphorene. As indicated in Fig. 1a, each Sn (Se) atom is bonded with three neighboring Se (Sn) atoms. The bond length is 2.73 Å ($R_{14}$, between atoms marked with 1 and 4) and 2.90 Å ($R_{12}$ and $R_{13}$), while the corresponding P-P bond length in phosphorene are 2.24 Å and 2.28 Å.[18] The bond angles formed by atoms 4, 1 and 2 or 3 are 90.8°, while the corresponding angles of phosphorene are much larger (103.5°),[18] leading to a much smaller 'armchair' opening for tinselenidene than phosphorene. For tinselenidene, the lattice parameters along *armchair* and *zigzag* directions are $a = 4.41$ Å and $b = 4.27$ Å, respectively, while for phosphorene $a = 4.58$ Å and $b = 3.32$ Å, exposing that the difference between $a$ and $b$ for tinselenidene is small, while for phosphorene it is drastic. The binding energy of tinselenidene with respect to bulk SnSe crystal is evaluated as $32\ \mathrm{meV}/\text{Å}^2$, which is larger than graphene ($17.8\ \mathrm{meV}/\text{Å}^2$) but close to phosphorene ($29.9\ \mathrm{meV}/\text{Å}^2$) (Fig. S3), revealing that tinselenidene is more difficult to be exfoliated than graphene, but can be synthesized using similar methods to phosphorene.

The phonon dispersion of tinselenidene is shown in Fig. 2a. No imaginary frequency is observed, indicating its kinetic stability. Interestingly, the phonon band profile show dramatically symmetry along $\Gamma - X$ and $\Gamma - Y$ directions. From the slope of longitudinal acoustic phonon branch at $\Gamma$ point, the sound speed (phonon group velocity) along $\Gamma - X$ (armchair) and $\Gamma - Y$ directions (zigzag) can be evaluated as 2.9 and 3.1 km/s, respectively, which are nearly isotropic and much slower than that of phosphorene (4.0 and 7.8 km/s).[17] The maximal frequency is 5.2 THz, only about 40% of phosphorene,[17] implying that tinselenidene is much softer than phosphorene. By means of phonon Boltzmann transport equation[32,33] and DFT, the LTC is calculated, as depicted in Fig. 2b. The LTC of tinselenidene at 300 K along *armchair* and *zigzag* directions are 2.20 and 2.54 $\mathrm{Wm^{-1}K^{-1}}$, respectively, illustrating a nearly isotropic phonon transport properties, which is much lower than and contrary to the anisotropic thermal lattice conductivity of phosphorene (14 and 30 $\mathrm{Wm^{-1}K^{-1}}$).[17] Among the known 2D materials that LTC had been studied,[34-37] tinselenidene may have the lowest LTC, implying its great potential for a good 2D thermoelectric material.

Figure 2c shows the electronic band structure of tinselenidene. Similar to the phonon band structure, the electronic structure also exhibits obvious symmetry along $\Gamma - X$ and $\Gamma - Y$ directions. The conduction band minimum (CBM) and valence band maximum (VBM) are marked with $C_X$, $C_Y$, $V_X$, $V_Y$, which locate at X' and Y' on $\Gamma - X$ and $\Gamma - Y$ lines (inset of Fig.

2c), respectively. At first glance, $C_X$ and $C_Y$, $V_X$ and $V_Y$ have the same energy, respectively. However, a closer inspection reveals a small but obvious deviation. $V_X$ is higher than $V_Y$ about 0.20 eV, and $C_Y$ is lower than $C_X$ about 0.04 eV, and therefore, $V_X$ is the valence band top (VBT) and $C_Y$ is the conduction band bottom (CBB), indicating that tinselenidene is a semiconductor with an indirect bandgap of 1.45 eV. Other calculation methods are also used to recheck this result, all of which support the observation of an indirect bandgap (Fig. S4). The direct gaps between $V_X$ and $C_X$, $V_Y$ and $C_Y$ are 1.49 and 1.65 eV, respectively. For a comparison, the bulk SnSe was found to be a semiconductor with an indirect gap of 1 eV [38] (0.923 eV)[39] and a direct gap of 1.2 eV[38] from the optical absorption measurements.

The effective mass of charge carriers can be extracted from the high-precise energy band calculation, as shown in Fig. 2d. The red solid and blue dash lines represent the effective mass of electron and hole, respectively, which are nearly a perfect circle, suggesting that the effective mass of them are nearly isotropic. As listed in Table I, the effective mass along $\Gamma - X$ and $\Gamma - Y$ are 0.14(e), 0.16(h) and 0.16(e), 0.18(h), respectively, indicating that effective mass of holes are slightly larger than that of electron. The small effective masses also suggest tinselenidene is likely to be a high carrier mobility 2D semiconductor. Note that the carrier effective mass of phosphorene[18] along $\Gamma - X$ and $\Gamma - Y$ are 0.17(e), 0.15(h) and 1.12(e), 6.35(h), respectively, showing an obvious anisotropy.

Despite of the isoelectronic and similar structure between tinselenidene and phosphorene, we have observed unexpected nearly symmetric phonon and electronic band structures and isotropic lattice thermal conductivity and effective mass of charge carriers, in sharp contrast to the strongly anisotropic properties of phosphorene. How do these unexpected properties origin from? Let us reexamine the geometrical structure of tinselenidene. Fig. 1d shows the top view of the tinselenidene, which reveals a character of square-like lattice with small deviation. In fact, tinselenidene can indeed be regarded as a distorted bilayer square 2D lattice, in which the upper sublayer is symmetric to the lower sublayer, and every Sn (Se) atom in upper sublayer is bonded with the Se (Sn) atom in lower sublayer. If we focus on all Se atoms in upper/lower sublayer, we would observe that they form obviously a square-like lattice. It is the same when we focus on all Sn atoms. Thus, the whole sublayer can be viewed as two square-like lattices nested within each other. Furthermore, the electron density (Fig. 1e) and the electrostatic potential (Fig. S1c) are extracted, which exhibit an obvious symmetry along *armchair* and *zigzag* directions, confirming that the electrons indeed move in square-like potential. The electron density around Se atoms is distinctly higher than that of Sn atoms, showing that electrons transfer from Sn atoms to Se atoms, which is consistent with the observation in projected density of states (Fig. S5). It may be owing to the large difference of electronegativity of Se and Sn atoms, implying that the Sn-Se bond may be polar covalent bonds with a strong polarity. Besides, a high electron density is found between the nonbonding neighboring Sn and Se atoms, revealing that strong nonbonding interactions exist between them, which are weaker than Sn-Se bond but may be still much stronger than common van der Waals interaction. Note that the nonbonding neighboring Sn-Se distance is 3.26 Å ($R_{15}$ and $R_{16}$), which is only 12% longer than the Sn-Se bond length (2.90 Å, $R_{12}$ and $R_{13}$), while in phosphorene the distance between nonbonding neighboring P atoms (3.41Å**)** is about 50% longer than P-P bond length (2.28 Å). The electronic localization function (ELF) of tinselenidene and phosphorene are also presented (Fig. S2). While the ELF of phosphorene shows typical obvious covalent bond character of P-P bond and is lack of interaction between nonbonding neighboring atoms, the ELF of tinselenidene clearly indicates the character of polar covalent bond and strong interactions between the nonbonding neighboring

Sn and Se atoms. From this point of view, the bonding properties of tinselenidene are different from that of phosphorene. Therefore, the geometrical structure of tinselenidene is in fact more symmetric than what we have seen in Fig. 1a, and an effective bilayer square-like 2D lattice emerges, which induces the symmetric phononic and electronic band structures and isotropic properties. It appears that the ball-stick model sometimes misleads our understanding on the geometrical structure and bonding nature of tinselenidene.

The strain effects on geometric, mechanical and electronic properties of tinselenidene are extensively studied. The geometric parameters of tinselenidene under uniaxial strains are shown in Fig. 3, in which $\varepsilon_x$, $\varepsilon_y$, and $\varepsilon_z$ indicate the relative strain along *x* (*armchair*), *y* (*zigzag*) and z directions, respectively, and the negative (positive) values represent compressive (tensile) strain. It is found that the layer thickness *c* (along *z* direction) increases with the increase of lattice parameter *a* (along *x* direction), giving rise to a negative Poisson's ratio of -0.17, showing that tinselenidene is a potential auxetic material. As the similar phenomenon was also observed in black phosphrous,[13,16] which may be common in materials with such hinge-like structures. However, the negative Poisson's ratio of tinselenidene emerges between armchair (*x*) and perpendicular *z* direction, while that appears between zigzag (*y*) direction and perpendicular *z* directions for phosphorene.[13] Besides, the absolute value of negative Poisson's ratio of tinselenidene is 6 times larger than phosphorene (-0.027).[13] Thus, the mechanism of negative Poisson's ratio in tinselenidene and phosphorene should be different, which may also be due to the strong nonbonding interaction in tinselenidene. Besides, a turning point in $\varepsilon_z - \varepsilon_x$ strain curve can be observed under 5% compressive strain along *x* direction, indicating a geometric phase transition, in which the space group of tinselenidene changes from Pmn21 (No. 31) to Pmmn (No. 59) (Fig. S6), behaving similarly with the geometric phase transition from Pnma (No. 62) to Cmcm (No. 63) in bulk SnSe under a high pressure.[40] The corresponding transition stress for tinselenidene is about 1.6 GPa, while the critical hydrostatic pressure for bulk is about 10.5 GPa.[40] Thus, a very low stress could induce the geometrical phase transition, which may be useful for manipulating the properties of tinselenidene. The stress-strain relations are given in the insets of Fig. 3, from which the Young's elastic modulus can be obtained as 24 GPa and 44 GPa along the *x* and *y* direction, respectively, showing a less anisotropic character than phosphorene (44 GPa and 166 GPa).[41] It is worth to note that tinselenidene is much more flexible than phosphorene and other isotropic 2D materials, such as graphene (1000 GPa),[42] h-BN (250 GPa),[43] and $MoS_2$ (330 GPa),[44] and may be the most flexible in the known 2D materials.[45]

Figure 4 presents the energy bands of tinselenidene under different uniaxial strains. The VBMs ($V_X$ and $V_Y$) and the CBMs ($C_X$ and $C_Y$) are indicated by small filled and unfilled squares and circles, respectively. When the compressive strain is increased along *x* direction, $C_X$ moves down and $V_Y$ moves up, leading to a shrinking indirect band gap, when the tensile strain is increased along *x* direction, $C_X$ moves up and $V_Y$ moves down, while $V_X$ and $C_Y$ are nearly fixed, leading to an intact indirect band gap. When the compressive strain is increased along *y* direction, $C_X$ moves up and $V_Y$ moves down, while $V_X$ and $C_Y$ are nearly fixed, leading to a nearly intact indirect band gap, when the tensile strain is increased along *y* direction, $C_X$ moves down and $V_Y$ moves up with the increase of the strain, leading to a shrinking indirect band gap. Fig. 4c and Fig. 4d show the strain effect on $C_X$, $C_Y$, $V_X$ and $V_Y$. One may note that the compressive strain along *y* direction affects the same way as the tensile strain along *x* direction, and vice versa. This is understandable, as the lattice parameters *a* and *b* are associated with positive Poisson’s ratio, when *a* is compressed, *b* is elongated, and vice versa. Another interesting fact is that $V_X$ is nearly always fixed whenever compressive and tensile strains are applied along *x* or *y* direction, $C_Y$ acts similarly when the tensile strain along *x* direction or the compressive strain along *y* direction is applied, $V_Y$ and $C_X$ are always sensitive and nearly linear response to any strain, showing a anisotropic response of electronic structures to the strain. From Fig. 4c and 4d, the colored areas marked by II and V correspond to the direct band gap, while I, III, IV and VI represent the indirect band gap. Thus, a strain-induced direct and indirect band gap transition can be observed. Notice that only a small compressive stress along *x* direction (about 0.24 GPa) or a tensile stress along *y* direction (about 0.45 GPa) will make tinselenidene transit from an indirect gap to a direct band gap, when these strains continue to increase more than about 2.5% (about 1.14 GPa), the energy gap shrinks and transits from a direct one into an indirect one again. 10% compressive strain along *x* direction (4.4 GPa) and tensile strain (4.5 GPa) along *y* direction would reduce the bandgap to 0.3 and 0.6 eV, respectively (Fig. S7). The compressive strain along *y* direction or the tensile strain along *x* direction almost does not affect the bandgap. Therefore, a low stress could give rise to diverse electronic structures, which makes tinselenidene an excellent 2D semiconductor material for the strain band engineering.[46,47] The strain effect on the effective mass of charge carrier is studied as well (Fig. S8). It is found that the compressive strain along *x* direction and the tensile strain along *y* direction will decrease the carrier effective mass, and the biaxial tensile strain will dramatically increase the effective mass and induce a noticeable anisotropy. The electronic structure of tinselenidene is indeed very sensitive to strain.

As mentioned above, the small effective masses of carrier may lead to a high carrier mobility of tinselenidene. Using the deformation potential theory,[48,49] the charge carrier mobility can be predicted from the carrier effective mass, deformation potential constant, and effective 2D elastic modulus. Similar methods have been used to predict carrier mobility of other 2D materials, such as graphyne,[50] graphdiyne,[51] $TiS_3$ monolayer,[52] $MoS_2$ monolayer[53] and phosphorene[18]. Deformation potential constants describe the scattering caused by electron-acoustic phonon interactions, and effective 2D elastic modulus can be viewed as the 2D form of Young's modulus (Table S3). As listed in Table 1, the highest hole mobility is 11520-14880 $cm^2V^{-1}S^{-1}$ along *x* direction, which is nearly one order higher than that along *y* direction (1050-1180 $cm^2V^{-1}S^{-1}$), while the electron mobility along *x* direction and *y* direction are 1200-1350 $cm^2V^{-1}S^{-1}$ and 990-1100 $cm^2V^{-1}S^{-1}$, respectively. Hence, tinselenidene will exhibit strong anisotropic *p*-type properties in electronic transport. Interestingly, although the carrier effective masses are isotropic, the carrier mobility is far from isotropic, especially for hole. The effective 2D elastic modulus along *x* direction ($C_{2D_x}$) is nearly half of that along *y* direction ($C_{2D_y}$), and the deformation potential constant along *x* ($E_{l_x}$) for hole is dramatically smaller than that along *y* direction ($E_{1_y}$), which is obviously the origin of anisotropic hole mobility. $E_{l_x}$ for hole is evaluated from the energy change of VBT ($V_X$) under proper strain along *x* direction. As can be seen in Fig. 4a, $V_X$ is nearly fixed under different strain along *x* direction, which is consistent with the extremely small $E_{l_x}$, implying that the behavior of energy band under different strain can be viewed as an indicator of high carrier mobility. Thus, besides the small effective carrier masses, an ultralow deformation potential is also responsible for the ultrahigh hole mobility. Since the deformation potential theory can consider only the scattering effect of longitudinal acoustical phonon, while other scatterings from optical phonons, impurities, etc. are not included, the experimentally observed carrier mobility may be not as high as the one calculated here. Note that a high hole mobility of order 10000 $cm^2V^{-1}S^{-1}$ for phosphorene[18] was also predicted with similar methods, while in experiments the mobility was reported as about 1000 $cm^2V^{-1}S^{-1}$ [7,24] or even up to 6000 $cm^2V^{-1}S^{-1}$ at low temperature,[25] it may still have potential to be raised by improving experimental condition and methods. Thus, tinselenidene can also be expected to be a high-mobility 2D atomic material in practice. Furthermore, although the bandgap is found to be indirect, by considering it is very easy to achieve indirect-direct transition by a rather low stress, one may see that tinselenidene could be a promising excellent 2D semiconductor for nanoelctronics and optoelectronics. Another interesting fact that should be addressed here is that while both of phosphorene and tinselenidene show *p*-type properties, and the hole mobility of phosphorene along *y* direction is dominant, the hole mobility of tinselenidene along *x* direction is dominant, which may be owing to different anisotropic electronic response to strain. This also implies the underlying mechanic and electronic differences between tinselenidene and phosphorene.

**Conclusions**

In summary, by means of extensive *ab initio* calculations, we find that tinselenidene is a semiconductor with an indirect bandgap of 1.45 eV, and has a ultralow lattice thermal conductivity smaller than 3 $Wm^{-1}K^{-1}$ at 300 K and a hole mobility as high as 11000 $cm^2V^{-1}S^{-1}$. In contrast to phosphorene, which is the isoelectronic and a similar structure partner to tinselenidene and has strongly anisotropic mechanical, electronic, and optical properties, we observe that tinselenidene has nearly symmetric phonon and electronic band structures, leading to nearly isotropic lattice thermal conductivity and charge carrier effective mass, which can be attributed to the effectively symmetric square-like bilayer lattice structure. The strain effect shows that the geometric, mechanic and electronic properties of tinselenidene are sensitive to the strain. A very low stress (1.6 GPa) along *x* direction can induce a geometrical phase transition. Besides, a uniaxial strain can shift the extremes of different energy valleys asynchronously, giving rise to an indirect-direct bandgap transition under a rather low stress (< 0.5 GPa).

Although the effective mass of charge carrier is isotropic, the carrier mobility is anisotropic, which can be attributed to the anisotropic response to strain. Furthermore, tinselenidene has a large negative Poisson's ratio, which indicates that it may be an auxetic material. The rich properties of tinselenidene suggest that it should be an excellent 2D material candidate for nanomechanics, thermoelectrics and optoelectronics.

## Methods

Most of the calculations are performed using Vienna *ab initio* simulation package (VASP)[54] with the generalized gradient approximation of Perdew-Burke-Ernzerhof (PBE)[55] for the exchange-correlation potential and a projector augmented wave (PAW)[56] method. The kinetic energy cutoff for plane wave functions is set to 700 eV and the energy convergence threshold is set as $10^{-5}$ eV. The Monkhorst-Pack k-mesh[57] of 15×15×1 is employed to sample the irreducible Brillouin zone. The shape and volume for each cell were fully optimized and the maximum force on each atom is less than 0.002 eV/Å. The optB88-vdW functional[58] is adopted to consider the van der Waals interactions. The modified Becke-Johnson (mBJ)[59] method is adopted to calculate electronic band structures. The phonon dispersion is calculated using PHONOPY package[60] with the finite displacement method. The lattice thermal conductivity is calculated using ShengBTE code.[32-33] The effective masses are derived from the band structure. By the deformation potential theory, the carrier mobility in 2D materials are calculated using the equation[48-53]

$$\mu_{2D} = \frac{e\hbar^3 C_{2D}}{k_B T m_e^* m_d (E_l)^2} \quad (1)$$

where $m_e^*$ is the effective mass for the conveyor direction, $m_d$ is the average effective mass defined by $m_d = \sqrt{m_x^* m_y^*}$ , $T$ represents the temperature that is taken as 300K, and $E_l$ is the deformation potential constant that contains the VBM for hole and the CBM for electron along the conveyor direction, expressed as $E_l = \Delta \mathrm{V}/(\Delta l/l_0)$ , where $l_0$ is the lattice constant along the conveyor direction, $\Delta l$ is the distortion of $l_0$, and $\Delta \mathrm{V}$ is the energy change of the band with proper strain (the step is set as 0.5%). $C_{2D}$ represents the effective 2D elastic modulus, which we calculate by using the following equation

$$C_{2D} = \frac{1}{S_0} \frac{\partial^2 E}{\partial (l/l_0)^2}\bigg|_{l=l_0} \tag{2}$$

where $E$ is the total energy after deformation, and $S$ is the lattice volume at equilibrium for a 2D system.

**Acknowledgements**

The authors thank Dr. Li-Zhi Zhang of University of Utah, Prof. Wei Ji of RUC and Prof. Zhen-Gang Zhu of UCAS for helpful discussions. All calculations are performed on Nebulae (DAWN6000) in National Supercomputing Center in Shenzhen and MagicCube (DAWN5000A) in Shanghai Supercomputer Center, China. This work is supported in part by the NSFC (Grant No. 11004239, No. 11474279), the MOST (Grant No. 2012CB932901 and No.2013CB933401) of China, and the fund from CAS.

**Author Contributions**

Q.B.Y conceived and designed the research. L.C.Z, G.Z.Q., and W.Z.F carried out the calculations. H.J.C. and Q.R.Z. contributed to the discussion and analysis of results. Q.B.Y., L.C.Z. and G.S. co-wrote the paper. All the authors participated in the discussions and reviewed the manuscript. G.S. supervised the whole project.

**Additional Information**.

The supplementary Information accompanies this paper at: http://www.nature.com/******

**Competing financial interests**: The authors declare no competing financial interests.

**Caption of Figure 1**

**Figure 1** Schematic structures of tinselenidene. (a) Perspective view, (b) side view and (d) top view. Yellow balls denote Se atoms and gray balls denote Sn atoms. The numbers from 1 to 6 label the neighboring Sn and Se atoms. The primitive cell of tinselenidene is indicated by dashed lines, where $a$ and $b$ denote the lattice parameter in the $x$ (armchair) and $y$ (zigzag) direction, respectively. The side view of phosphorene (c) is included to compare with tinselenidene. The two-dimensional charge density of tinselenidene (e) is illustrated in the $xy$ plane crossing all Sn atoms. The red and blue colors depict the high and low charge density, respectively. The unit of charge density is e/bohr$^3$.

**Caption of Figure 2**

**Figure 2** (a) The phonon dispersion for tinselenidene, where three acoustic phonon branches are indicated as LA, TA and ZA. (b) The lattice thermal conductivity of tinselenidene in the armchair ($x$) and zigzag ($y$) directions, respectively. (c) The band structure of tinselenidene, where black lines represent the calculations with mBJ functional and red lines with HSE06 method. The CBM and VBM for the band structure are marked as $C_Y$, $C_X$ and $V_Y$, $V_X$, respectively. (d) The effective mass for electrons and holes in tinselenidene.

**Caption of Figure 3**

**Figure 3** The mechanical response of tinselenidene under uniaxial strain along $x$ (a) and $y$ (b) directions. Strain is defined as $s = (l - l_0)/l_0$, where $l = a, b, c$ represent the lattice parameters (thickness for $c$) along $x, y, z$ directions under strain, respectively, and $l_0 = a_0,\ b_0,\ c_0$ are the corresponding original lattice constants (thickness for $c_0$) without strain. The positive (or negative) s means a tensile (or compressive) strain, while $s = 0$ corresponds the case without strain. The Poisson's ratio can be obtained by fitting $s = -\nu_1 t + \nu_2 t^2 + \nu_3 t^3$, where $t$ is the strain along the $x$ or $y$ direction, and $\nu_1$ could be regarded as the Poisson's ratio. The corresponding stress-strain relations in $x$ and $y$ directions are shown in the upper right insets.

**Caption of Figure 4**

**Figure 4** Band structure of tinselenidene calculated by mBJ method under uniaxial strain from -10% to 10% along $x$ (a) and $y$ (b) directions. The CBM and VBM in the band structure are marked by $C_Y$, $C_X$ and $V_Y$, $V_X$, respectively. The energies of $C_Y$, $C_X$, $V_Y$ and $V_X$ as a function of strain in the *x* (c) and *y* (d) directions are plotted. The indirect band gap appears in regions I, III, IV and VI, and the direct band gap appears in the regions II and V. The transition from an indirect gap to a direct gap can be determined by the energy crossover.

**TABLES**.

**Table 1.** The effective mass and mobility of charge carriers in tinselenidene

| Carrier type | $m_x^*/m_0$ | $m_y^*/m_0$ | $E_{l_x}$ | $E_{l_y}$ | $C_{2D_x}$ | $C_{2D_y}$ | $\mu_{2D_x}$ | $\mu_{2D_y}$ |
|---|---|---|---|---|---|---|---|---|
| | Γ-X | Γ-Y | (eV) | | ($Jm^{-2}$) | | ($10^3 cm^2V^{-1}S^{-1}$) | |
| electron | 0.143 | 0.158 | -3.28±0.10 | 4.65±0.13 | 13.8 | 25.1 | 1.20-1.35 | 0.99-1.10 |
| hole | 0.155 | 0.175 | -0.94±0.06 | -4.09±0.12 | 13.8 | 25.1 | 11.52-14.88 | 1.05-1.18 |

**Table 1** The predicted effective mass and mobility of carriers in tinselenidene. $m_x^*$ and $m_y^*$ are effective mass along Γ-X and Γ-Y directions, respectively. $E_{l_x}$ ($E_{l_y}$) and $C_{2D_x}$ ($C_{2D_y}$) represent the deformation potential constant and effective 2D elastic modulus for *x* (*y*) direction. $C_{2D_x}$ and $C_{2D_y}$ denote the carrier mobility in *x* and *y* direction, respectively. Note that the calculated effective 2D elastic modulus is consistent with the Young's modulus evaluated from stress-strain relations (see Table S3).